\documentclass[a4paper,fleqn]{cas-sc}

\usepackage[numbers,compress]{natbib}

\def\tsc#1{\csdef{#1}{\textsc{\lowercase{#1}}\xspace}}
\tsc{WGM}
\tsc{QE}

\begin{document}
\let\WriteBookmarks\relax
\def\floatpagepagefraction{1}
\def\textpagefraction{.001}

\shorttitle{}    

\shortauthors{}  

\title [mode = title]{Even--odd classification of reentrant topology in a Su--Schrieffer--Heeger chain with integer-power quasiperiodic hopping}  


\author[1]{Yusheng Niu}
\credit{Writing -- original draft, Formal analysis, Data curation}

\author[2]{Hui Liu}
\cormark[1]
\ead{liuhui99@sxu.edu.cn}
\credit{Writing -- original draft, Investigation}

\author[2]{Zhihao Xu}
\cormark[1]
\ead{xuzhihao@sxu.edu.cn}
\credit{Writing -- review \& editing, Supervision, Investigation, Funding acquisition}

\affiliation[1]{organization={Sanli Honors College, School of Physics and Electronics Engineering, Shanxi University},
            city={Taiyuan},
            postcode={030006}, 
            country={China}}

\affiliation[2]{organization={Institute of Theoretical Physics and State Key Laboratory of Quantum Optics Technologies and Devices, Shanxi University},
            city={Taiyuan},
            postcode={030006}, 
            country={China}}

\cortext[1]{Corresponding author}



\begin{abstract}
We investigate reentrant topology in a one-dimensional Su--Schrieffer--Heeger chain with integer-power quasiperiodically modulated intracell hopping. The modulation is controlled by a positive integer exponent $n$ and a tunable parameter $\beta$, which interpolates between a smooth integer-power quasiperiodic profile and a sign-function limit. Combining the zero-mode inverse-localization-length criterion with a real-space topological indicator, we determine the phase diagrams in the $\beta\to0$, $\beta\to\infty$, and finite-$\beta$ regimes. We find an even--odd classification of reentrant topological windows governed by the dc component and support of the full modulation profile. For positive modulation strength, odd powers yield a zero-mean sign-changing profile and can induce reentrance from the clean trivial regime $|t_1|>1$, whereas even powers yield a positive-mean non-negative profile and allow reentrance only from the negative clean trivial regime $t_1<-1$. Although even-power profiles contain sign-changing fluctuations after subtracting their positive average, the full modulation profile remains non-negative. This even--odd structure of the full modulation profile provides a control knob for topological Anderson-like reentrant phases. We further discuss the associated bulk localization properties and show that the phase diagrams are robust against moderate hopping fluctuations, suggesting a feasible realization in electrical circuits.
\end{abstract}




\maketitle

\section{Introduction}

Topological phases of matter have attracted sustained interest because of their robust boundary states and unconventional transport
properties~\cite{PRB2011Fulga,PRL2013Cai,PRB2021Longhi,RevModPhys2016Bansil,PRL2008Wang,Nature2008Wang,PRA2025Sinha,LIMA2026131574}. In one-dimensional systems, the Su--Schrieffer--Heeger (SSH) model serves as a
paradigmatic platform for studying chiral-symmetry-protected topological phases~\cite{PRA2025Sinha,PRB2024Cinnirella,PRB2023Nava,PRB2020Scollon,FrontiersZhanpeng2025}. In its topological regime, the SSH chain supports
zero-energy edge states under open boundary conditions, and its phase transition can be characterized by bulk topological invariants~\cite{PRL2013Cai,Ann2022Lu,PRA2022Tang,PRB2024Zuo,JP2026Kumar} as well as by the
localization properties of zero modes~\cite{Ann2022Lu,PRA2022Tang,PRB2024Zuo,JP2026Kumar}.

Disorder and aperiodic modulations can significantly modify the topological properties of low-dimensional
systems~\cite{PRA2025Sinha,PRB2024Assun,PRB2024Ouyang,PRB2021Rai,PRB2024Kobia,PRB2024Sandberg,CP2025Ji,PRB2024Padhan}. While disorder generally tends to localize quantum states~\cite{PRB2017Mansha,PRL2017Liu}, it may also
induce nontrivial topological phases, leading to the so-called topological Anderson insulator
(TAI)~\cite{FrontiersZhanpeng2025,Ann2022Lu,PRA2022Tang,PRB2024Zuo,PRB2021Lin,PRB2021Zhang,PRL2009Li,PRL2009Groth,PRB2009Jiang,PRB2023Liu,Science2018Meier,PRL2014Mondragon,PRB2023Zhang,PRB2011Xing,PRB2012Zhang,PRL2015Titum,SciRep2016Orth,PRB2019Zhang,PRL2024Zhao,PRL2020Liu,NC2022Lin,PRL2022Wang,PRL2022Cui,PRL2024Ren,PRResearch2025Li,PRResearch2021Bagrets,PRL2025Grindall,PRResearch2024Li,Opt2020Longhi}.
Here the term TAI is used in a broader sense, referring also to reentrant topological phases induced by deterministic quasiperiodic modulations~\cite{FrontiersZhanpeng2025,PRB2024Padhan,Ann2022Lu}, in analogy with the
disorder-driven TAI originally proposed for random systems~\cite{PRL2009Li,PRL2009Groth,PRB2011Xing,PRB2012Zhang,SciRep2016Orth}. We emphasize that, since the modulation considered in this work is deterministic rather than
random, the term ``TAI'' should be understood as a topological Anderson-like phase in a generalized sense. In particular, the essential feature discussed below is the reentrant transition from a clean trivial phase to a
nontrivial topological phase induced by a spatial modulation~\cite{PRB2025Deng,PRB2024Ghosh}, rather than a conventional disorder-driven Anderson mechanism. 

In parallel, quasiperiodic systems provide an intermediate setting between periodic order and random disorder, and have become an important platform for exploring the interplay among
topology~\cite{PRA2025Sinha,PRB2024Ouyang,PRB2021Rai,PRB2024Kobia,PRB2024Sandberg,CP2025Ji,PRB2024Padhan},
localization~\cite{PRB2024Sandberg,PRB2024Padhan,PRB2023Dai,PRB2024Wang,PRL2020Sbroscia,wang2025powerlaw,PRL2021Gautier,PRL2023Zhu,PRA2024Zhu,SciPostPhys2022Tong}, and spectral
criticality~\cite{PRB2023Dai,PRL2020Yao,RevModPhys2021Jagannathan}. In recent years, disorder- or quasiperiodicity-driven topological transitions have been extensively investigated in SSH-type lattices and related
one-dimensional models~\cite{PRB2021Longhi,PRB2020Scollon,FrontiersZhanpeng2025,Ann2022Lu,PRA2022Tang,PRB2019Longhi,PRL2014Altland}. Quasiperiodically modulated SSH-type chains have already been shown to support
topological Anderson-like phases and reentrant topological behavior. For example, Ref.~\cite{FrontiersZhanpeng2025} studied a generalized SSH quasicrystal and revealed TAI-like and reentrant TAI-like phases, together with
their connection to reentrant localization transitions. Ref.~\cite{Ann2022Lu} demonstrated exact mobility edges and a topological Anderson-like insulating phase in a slowly varying quasiperiodic model.
Ref.~\cite{PRA2022Tang} further showed that quasiperiodic SSH chains can host TAI-like phases with extended, intermediate, and localized bulk states. These works established that quasiperiodic hopping modulation can induce
nontrivial topology and that the resulting TAI-like phases are not necessarily tied to a conventional Anderson localization transition.

The present work addresses a different question. Instead of asking whether a quasiperiodic SSH chain can host TAI-like or reentrant TAI-like phases, we focus on how the internal profile of the quasiperiodic hopping modulation controls the formation of reentrant topological windows. The integer-power modulation considered here provides a minimal setting for this purpose. Odd powers of $\cos x$ generate a zero-mean full modulation profile that changes sign about zero, whereas even powers generate a positive-mean non-negative full modulation profile for positive $\lambda$. An even-power profile can also be decomposed into this positive average and a zero-mean fluctuation that changes sign. Therefore, the relevant even--odd distinction is not the absence of sign-changing fluctuations for even $n$, but whether the full modulation profile itself is sign-changing or non-negative. For positive $\lambda$, odd powers can induce reentrant topology from the clean trivial regime $|t_1|>1$, while even powers can do so only from the negative trivial regime $t_1<-1$. Thus, the novelty of the present work lies not in the mere observation of a quasiperiodicity-induced reentrant topological phase, but in identifying the dc component and support of integer-power quasiperiodic hopping profiles as control knobs that produce an even--odd classification of reentrant topological windows.

Motivated by this issue, we study a generalized SSH chain with a tunable hopping modulation~\cite{PRB2023Dai,wang2025powerlaw}, where a control parameter $\beta$ continuously interpolates between the smooth
integer-power quasiperiodic limit and a sign-function limit that becomes square-wave-like for odd $n$ and effectively uniform for even $n$. This unified setting enables us to examine not only the limiting
cases but also the continuous evolution of the topological phase diagram between them.

The $\tanh$-regulated profile in Eq.~(\ref{eq2}) can be viewed as a bounded or saturated integer-power quasiperiodic hopping modulation. Beyond electrical circuits, related bounded hopping profiles may be engineered
in photonic waveguide arrays through designed waveguide spacings or in cold-atom momentum lattices through tunable Raman couplings~\cite{PRL2020Xie,PRL2022majie}. Nonlinear or saturable photonic systems may also
approximate tanh-like responses, while the circuit platform discussed below offers a direct way to realize tunable quasiperiodic hoppings, including effective positive and negative couplings.

In this work, we combine analytical and numerical methods to investigate the topological phase transitions in this model. The central finding is an even--odd classification of the phase diagrams with respect to the integer
exponent $n$: odd and even exponents lead to qualitatively different reentrant windows because they generate different dc components and supports of the full modulation profile. By analyzing the zero-mode inverse
localization length, we obtain exact analytical expressions for $n=1,2,3,4$, which yield explicit or implicit transition conditions. These analytical results are confirmed by numerical evaluations of the real-space
topological indicator~\cite{PRB2011Fulga,PRB2024Zuo,Opt2020Longhi}. We show that deterministic quasiperiodic modulation can induce TAI-like reentrant topological phases within finite parameter windows. We also propose an
electrical-circuit realization of the model, which offers a possible route for experimental verification. 

The remainder of this paper is organized as follows. In Section \ref{sec:model}, we introduce the model and the topological characterization methods. In Section \ref{sec:phase}, we present the topological phase diagrams in three parameter regimes ($\beta\to 0$, $\beta\to\infty$, and finite $\beta$), consolidate the results into a unified picture of even--odd reentrant topology, and briefly discuss the relation between the topological phase diagrams and bulk localization properties. In Section \ref{sec:experiment}, we outline an electrical-circuit implementation. Section \ref{sec:conclusion} summarizes the main findings. Analytical details are provided in Appendix \ref{sec:calculation}. Supplementary discussion on localization properties is presented in Appendix \ref{sec:bulk}. Robustness analysis of the topological phase diagram with respect to experimental component errors is given in Appendix \ref{sec:robustness}.

\section{Model and topological criterion}\label{sec:model}

\begin{figure}
\centering
\includegraphics[width=0.65\linewidth]{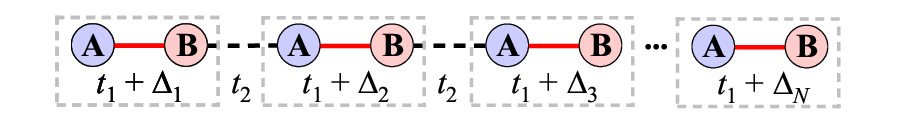}
\caption{Schematic illustration of the SSH chain with quasiperiodically modulated intracell hopping $\omega_j=t_1+\Delta_j$ and uniform intercell hopping $t_2$.\label{Fig1}}
\end{figure}

We consider a one-dimensional SSH chain with $L=2N$ lattice sites, corresponding to $N$ unit cells, as illustrated in Fig.~\ref{Fig1}. Each unit cell contains two sublattices, labeled $A$ and $B$. The red solid lines denote the intracell hopping, whereas the black dashed lines denote the intercell hopping. Within the tight-binding approximation, the Hamiltonian is given by
\begin{equation}
\label{eq1}
H=\sum_{j}\omega_j\left(c_{j,B}^{\dagger}c_{j,A}+\mathrm{h.c.}\right)+t_2\sum_{j}\left(c_{j+1,A}^{\dagger}c_{j,B}+\mathrm{h.c.}\right),
\end{equation}
where $\omega_j=t_1+\Delta_j$ is the intracell hopping amplitude, with $t_1$ the uniform part and $\Delta_j$ the quasiperiodic modulation. The intercell hopping amplitude $t_2$ is taken as the energy unit, i.e.,~$t_2=1$. Here, $c_{j,A}$ ($c_{j,B}$) and $c_{j,A}^{\dagger}$ ($c_{j,B}^{\dagger}$) denote the annihilation and creation operators on sublattice $A$ ($B$) in the $j$th unit cell, respectively. The modulation is chosen as
\begin{equation}
\label{eq2}
\Delta_j=\lambda \frac{\tanh\!\left[\beta \cos^n(2\pi \alpha j)\right]}{\tanh \beta},
\end{equation}
where $\lambda$ denotes the modulation strength and $\beta$ is a tunable parameter controlling the profile of the modulation. The exponent $n$ is taken to be a positive integer, and $\alpha=(\sqrt{5}-1)/2$ is taken as the irrational modulation frequency.

This modulation continuously interpolates between different limiting cases. In the limit $\beta\to 0$, one has $\Delta_j\approx \lambda \cos^n(2\pi \alpha j)$, which corresponds to a smooth integer-power quasiperiodic modulation. In the opposite limit $\beta\to \infty$, the modulation approaches $\Delta_j\approx \lambda\,\mathrm{sgn}[\cos^n(2\pi \alpha j)]$. For odd $n$, one has $\mathrm{sgn}[\cos^n x]=\mathrm{sgn}(\cos x)$, so that the modulation becomes square-wave-like. For even $n$, $\mathrm{sgn}[\cos^n(2\pi\alpha j)]=1$ for all lattice sites, so that the modulation becomes effectively uniform. In this work, we investigate the topological phase transitions induced by this quasiperiodic modulation in different $\beta$ regimes.

Because the Hamiltonian in Eq.~(\ref{eq1}) contains only off-diagonal hopping terms, the model preserves chiral symmetry. For this class of one-dimensional chiral systems, the topological phase boundary can be determined
from the zero-energy transfer equation. At zero energy, the eigenvalue equation decouples into two independent recurrence relations on the two sublattices~\cite{PRA2022Tang}. For the $A$ sublattice, one obtains
\begin{equation}
\psi_{j+1,A}=-\frac{\omega_j}{t_2}\psi_{j,A},
\end{equation}
with an analogous relation for the $B$ sublattice. The inverse localization length of the zero-energy mode is determined by the exponential growth or decay rate of the corresponding transfer
product~\cite{Ann2022Lu,PRA2022Tang,PRB2024Zuo}. We define it as
\begin{equation}
\label{eq3}
\nu=\left|\lim_{N\to\infty}\frac{1}{N}\sum_{j=1}^{N}\ln\left|\frac{t_2}{\omega_j}\right|\right|.
\end{equation}
Here $\nu$ is non-negative by definition. Therefore, it characterizes the localization length of the zero-energy mode and determines the phase boundary through
$\nu=0$~\cite{PRA2025Sinha,Ann2022Lu,PRA2022Tang,PRB2024Zuo,JP2026Kumar}. In the following, we use $\nu=0$ to determine the analytical phase boundaries, while the topological character of each region is identified
by the real-space topological indicator.

For chiral SSH-type chains with off-diagonal modulation, a convenient real-space topological indicator is~\cite{PRB2011Fulga,PRB2024Zuo,JP2026Kumar,Opt2020Longhi}
\begin{equation}
\label{eq4}
Q=\frac{1}{2}\left[1-\mathrm{sgn}\left(\prod_j \omega_j^2-\prod_j t_2^2\right)\right],
\end{equation}
where the product runs over all unit cells. Specifically, $Q=1$ and $Q=0$ correspond to the topologically nontrivial and trivial phases, respectively. The phase boundary obtained from $Q$ is consistent with the zero-mode condition $\nu=0$. In the following, we use $Q$ to construct the numerical phase diagrams and $\nu=0$ to derive the analytical phase boundaries. In the figures, the topologically nontrivial and trivial regions are indicated by purple and gold, respectively.

In the numerical calculations, the quasiperiodic modulation is defined on the unit-cell index $j$. We therefore choose the number of unit cells to be a Fibonacci number, $N=610$, corresponding to $L=2N=1220$ lattice sites, unless otherwise specified. We have also checked larger Fibonacci sizes, such as $N=2584$ and $N=6765$ in the localization analysis, and found that the topological phase boundaries extracted from $Q$ converge to the analytical boundaries.

\section{Phase diagrams}\label{sec:phase}

Unless otherwise stated, we focus on positive modulation strength $\lambda>0$ in the phase diagrams. For odd $n$, the phase diagram is symmetric under $\lambda\to-\lambda$ because the full modulation profile is sign-changing and zero-mean. For even $n$, the profile itself is non-negative; changing the sign of $\lambda$ reverses the direction of the induced hopping shift and correspondingly exchanges the roles of the positive- and negative-$t_1$ sides of the clean trivial regime. Throughout this section, and in the main text more generally, we focus on the topological phase diagrams; the bulk localization properties are discussed separately in Appendix \ref{sec:bulk}.

\subsection{$\beta \to 0$ limit}\label{sec:beta0}

We first consider the limit $\beta\to 0$, in which the intracell hopping reduces to $\omega_j=t_1+\lambda \cos^n(2\pi\alpha j)$. Setting $t_2=1$ in Eq.~(\ref{eq3}), the inverse localization length of the zero-energy mode can be written as
\begin{equation}
\label{eq5}
\nu=\left|\lim_{N\to\infty}\frac{1}{N}\sum_{j=1}^{N}\ln\left|t_1+\lambda \cos^n\!\left[2\pi\alpha (j+j_0)\right]\right|\right|,
\end{equation}
where $j_0$ denotes an arbitrary initial site index. Because $\alpha$ is irrational, the sequence $2\pi\alpha(j+j_0)$ is equidistributed modulo $2\pi$ in the thermodynamic limit. The spatial average in
Eq.~(\ref{eq5}) can therefore be replaced by an integral over one period~\cite{PRB2019Longhi,PRB2021Longhi}, yielding
\begin{equation}
\label{eq6}
\nu=\left|\frac{1}{2\pi}\int_{0}^{2\pi}\ln\left|t_1+\lambda \cos^n x\right|dx\right|.
\end{equation}
The topological phase boundary is determined by the zero-mode delocalization condition $\nu=0$.

\begin{figure}
\includegraphics[width=\linewidth]{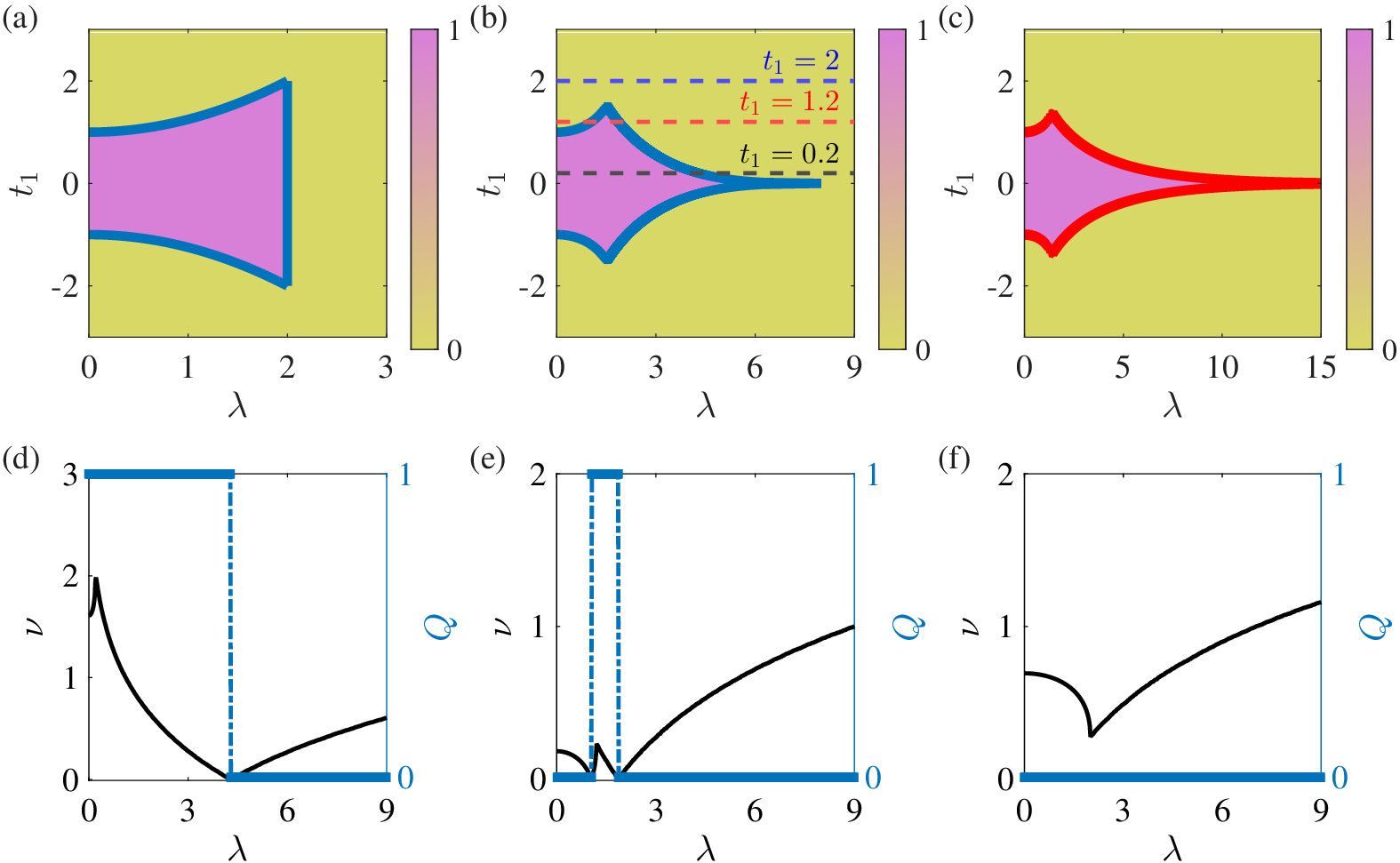}
\caption{\label{Fig2}Topological phase diagrams and zero-mode inverse-localization-length analysis in the $\beta\to 0$ limit for odd $n$. The colored regions in (a)--(c) are obtained from the numerical topological indicator $Q$, while the solid lines denote the phase boundaries determined from the criterion $\nu=0$. The blue solid lines represent analytical phase boundaries, and the red solid lines denote the boundaries obtained from numerical integration of Eq.~(\ref{eq6}). (a) $n=1$. (b) $n=3$, where the black, red, and blue horizontal dashed lines indicate $t_1=0.2$, 1.2, and 2, respectively, corresponding to the cuts shown in (d), (e), and (f). (c) $n=5$. (d) $n=3$, $t_1=0.2$. (e) $n=3$, $t_1=1.2$. (f) $n=3$, $t_1=2$. Here $N=610$ unit cells are used, corresponding to $L=1220$ lattice sites.}
\end{figure}

For odd $n$,  Figs.~\ref{Fig2}(a)--\ref{Fig2}(c) show the phase diagrams in the $(t_1,\lambda)$ plane for $n=1$, 3, and 5, respectively. In all odd-$n$ cases considered here, the system can enter a TAI-like reentrant topological phase within a finite interval of modulation strength for an appropriate range of $t_1$.

For $n=1$, the analytical phase boundaries derived in Refs.~\cite{Ann2022Lu,PRA2022Tang} are
\begin{equation}
\label{eq7}
\left\{
\begin{aligned}
& |t_1|=\frac{\lambda^2}{4}+1,&&
\qquad |t_1/\lambda|>1,\qquad |\lambda|\leq 2, \\
& |\lambda|=2,&&
\qquad 0<|t_1/\lambda|<1.
\end{aligned}
\right.
\end{equation}
These analytical boundaries, plotted as blue solid curves in Fig.~\ref{Fig2}(a), agree very well with the numerical phase diagram. The two branches of Eq.~(\ref{eq7}) merge at the terminal point $(\lambda, |t_1|)=(2,2)$, beyond which the first branch ceases to exist. In particular, within the interval $1<|t_1|<2$ (where the upper bound is set by this terminal point), the system undergoes a reentrant transition: it is trivial at weak modulation, enters a TAI-like topological phase at intermediate $\lambda$, and returns to the trivial phase when $\lambda$ is further increased.

For $n=3$, the zero-mode inverse localization length can likewise be evaluated analytically. The detailed derivation is presented in Appendix \ref{appendix1}, where the integral in Eq.~(\ref{eq6}) is reduced to an exact analytic expression. The phase boundary then follows from the condition $\nu=0$ applied to that expression. The resulting analytical boundary is plotted in Fig.~\ref{Fig2}(b), showing good agreement with the numerical results. In the interval approximately given by $1<|t_1|\lesssim 1.5$, where the upper bound is obtained from the condition $\nu=0$ using the exact expression in Appendix \ref{appendix1}, the system also exhibits a reentrant transition sequence from a trivial phase to a TAI-like reentrant topological phase and then back to the trivial phase as $\lambda$ increases.  Figs.~\ref{Fig2}(d)--\ref{Fig2}(f) further display $\nu$ and $Q$ as functions of $\lambda$ for representative values of $t_1$. For $t_1=0.2$, the system undergoes a single topological-to-trivial transition, signaled simultaneously by a jump in $Q$ and the vanishing of $\nu$. For $t_1=1.2$, two transition points appear, confirming the emergence of a TAI-like reentrant topological phase. By contrast, for $t_1=2$, the system remains trivial for all $\lambda$.

The same qualitative behavior persists for higher odd exponents. For example, the phase boundary shown for $n=5$ in Fig.~\ref{Fig2}(c) is obtained numerically from the zero-mode inverse-localization-length criterion in Eq.~(\ref{eq6}), and displays the same reentrant structure. This indicates that the TAI-like reentrant topological phase persists for odd $n\geq 3$.

The odd-$n$ results show that, for positive $\lambda$, a zero-mean sign-changing full modulation profile provides both positive and negative corrections to the bare hopping and can induce a TAI-like reentrant phase from the clean trivial regime $|t_1|>1$, although the actual existence range and width of this window depend on $n$. Thus, the condition $|t_1|>1$ should be understood as the clean trivial regime from which the reentrant phase can be induced, rather than as a sufficient condition for its existence at arbitrary modulation strength.

\begin{figure}
\includegraphics[width=\linewidth]{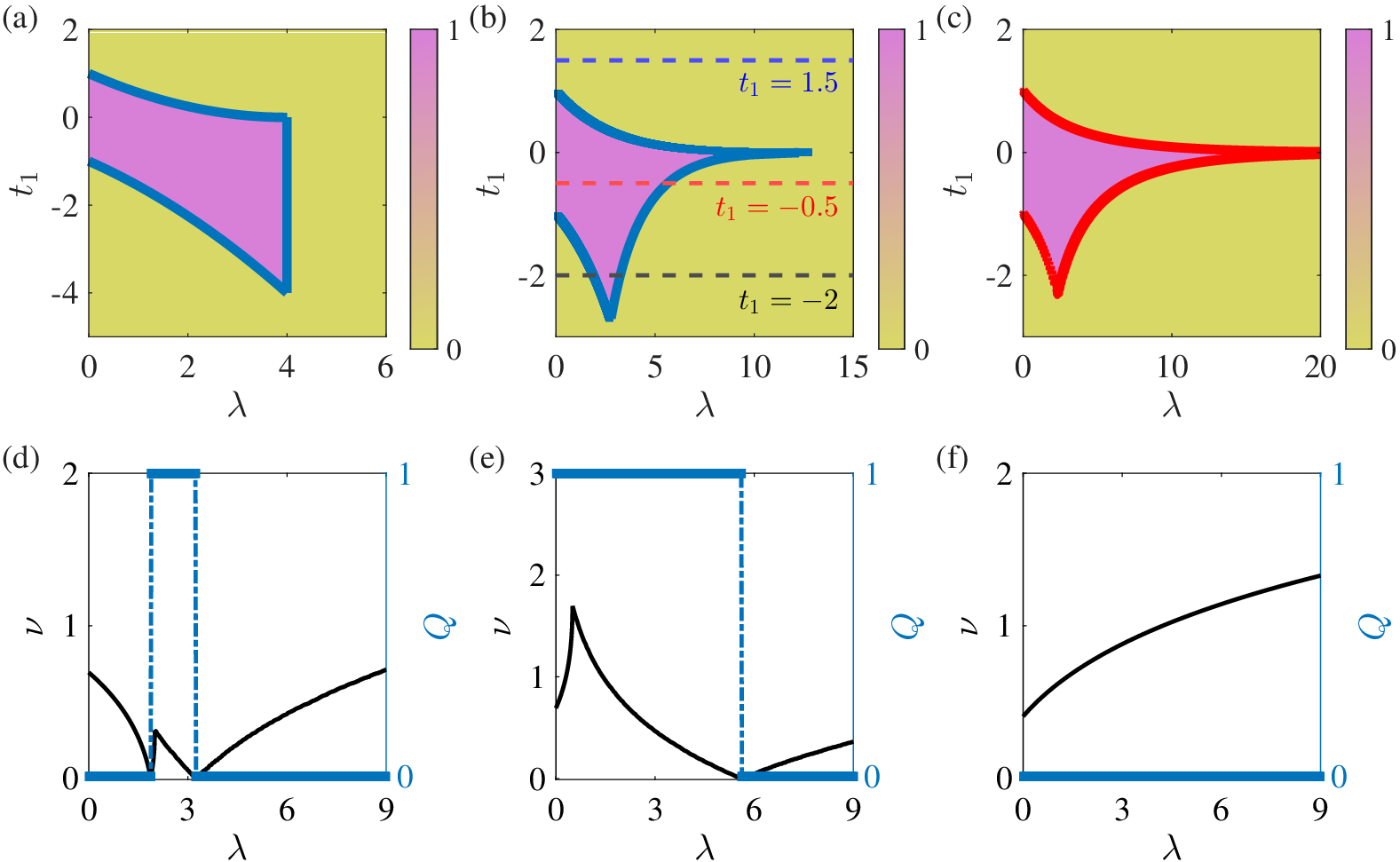}
\caption{\label{Fig3}Topological phase diagrams and zero-mode inverse-localization-length analysis in the $\beta\to 0$ limit for even $n$. The colored regions in (a)--(c) are obtained from the numerical topological indicator $Q$, while the solid lines denote the phase boundaries determined from the criterion $\nu=0$. The blue solid lines represent analytical phase boundaries, and the red solid lines denote the boundaries obtained from numerical integration of Eq.~(\ref{eq6}). (a) $n=2$. (b) $n=4$, where the black, red, and blue horizontal dashed lines indicate $t_1=-2$, $-0.5$, and $1.5$, respectively, corresponding to the cuts shown in (d), (e), and (f). (c) $n=6$. (d) $n=4$, $t_1=-2$. (e) $n=4$, $t_1=-0.5$. (f) $n=4$, $t_1=1.5$. Here $N=610$ unit cells are used, corresponding to $L=1220$ lattice sites.}
\end{figure}

In contrast to the odd-$n$ case, the even-$n$ phase diagrams shown in  Figs.~\ref{Fig3}(a)--\ref{Fig3}(c) display a qualitatively different structure for $n=2$, 4, and 6. The system again enters a TAI-like reentrant topological phase within a finite interval of $\lambda$ for suitable $t_1$, but the relevant parameter region is now restricted to negative $t_1$. For $n=2$, analytical evaluation of Eq.~(\ref{eq6}) yields
\begin{equation}
\label{eq10}
\nu=
\left\{
\begin{aligned}
& \left|2\ln \left| 1+\sqrt{\lambda/t_1+1} \right|+\ln |t_1|-\ln 4\right|,&& \qquad \lambda/t_1>-1, \\
& \left|\ln \left| \lambda/4 \right|\right|,&& \qquad \lambda/t_1<-1,
\end{aligned}
\right.
\end{equation}
from which the phase boundaries follow as
\begin{equation}
\label{eq11}
\left\{
\begin{aligned}
& t_1=1-\frac{\lambda}{2}+\frac{\lambda^2}{16}
\quad\text{or}\quad
t_1=-\!\left(1+\frac{\lambda}{2}+\frac{\lambda^2}{16}\right),&&
\qquad \lambda/t_1>-1, \\
& |\lambda|=4,&&
\qquad \lambda/t_1<-1.
\end{aligned}
\right.
\end{equation}
The detailed derivation of $\nu$ is presented in Appendix \ref{appendix2}. A useful interpretation of Eq.~(\ref{eq11}) follows from $\lambda\cos^2x=\lambda/2+(\lambda/2)\cos2x$. Thus, for $n=2$,
$\omega_j=\widetilde t_1+\widetilde\lambda\cos(2\pi\widetilde\alpha j)$,
where $\widetilde t_1=t_1+\lambda/2$, $\widetilde\lambda=\lambda/2$, and
$\widetilde\alpha=2\alpha$. Hence, in the $\beta\to0$ limit, the $n=2$ model is equivalent to a conventional single-cosine-modulated SSH chain along a $\lambda$-dependent path in the effective parameter space; substituting these effective parameters into the $n=1$ boundary in Eq.~(\ref{eq7}) reproduces Eq.~(\ref{eq11}). This mapping also explains the asymmetry in the original variables. Although the fluctuating part $(\lambda/2)\cos2x$ changes sign around the positive average $\lambda/2$, the full modulation profile remains non-negative, $0\leq\lambda\cos^2x\leq\lambda$ for $\lambda>0$. At fixed bare $t_1$, increasing $\lambda$ simultaneously increases both the effective hopping $\widetilde t_1$ and the fluctuation amplitude $\widetilde\lambda$. Therefore, the positive shift can compensate a negative trivial hopping $t_1<-1$, but cannot reduce a positive trivial hopping $t_1>1$, explaining why the reentrant TAI-like phase for even $n$ and positive $\lambda$ is confined to the negative-$t_1$ side. The analytical boundaries shown in Fig.~\ref{Fig3}(a) agree very well with the numerical phase diagram. In the interval $-4<t_1<-1$, the system undergoes a reentrant transition from the trivial phase to a TAI-like reentrant topological phase and finally back to the trivial phase as $\lambda$ increases. For $n=4$, the zero-mode inverse localization length can likewise be evaluated analytically, and its exact analytic expression is presented in Appendix \ref{appendix3}. The corresponding phase boundaries determined from $\nu=0$ are
\begin{equation}
\label{eq13}
\left\{
\begin{aligned}
& \left| 1+\sqrt{1+\kappa} \right|^2 \left| 1+\sqrt{1-\kappa} \right|^2 |t_1|=16,&&
\qquad \lambda/t_1>-1, \\
& \left| 1+\sqrt{1+\kappa} \right|^2 |\kappa t_1|=16,&&
\qquad \lambda/t_1<-1,
\end{aligned}
\right.
\end{equation}
where $\kappa=\sqrt{-\lambda/t_1}$, with the principal branch understood when $\kappa$ becomes complex. These analytical boundaries, plotted in Fig.~\ref{Fig3}(b), again agree very well with the numerical results. In the interval $-2.74<t_1<-1$ (where the lower bound is determined numerically from the condition $\nu=0$ using the exact analytic expression for the zero-mode inverse localization length in Appendix \ref{appendix3}), the system exhibits the same reentrant behavior. For higher even exponents such as $n=6$, the phase boundary shown in Fig.~\ref{Fig3}(c) is likewise obtained numerically from the zero-mode inverse-localization-length criterion in Eq.~(\ref{eq6}). The resulting phase diagram further confirms that this behavior remains qualitatively unchanged for higher even exponents, suggesting that the TAI-like reentrant topological phase is a robust feature for even $n\geq 4$. Unlike the special $n=2$ case, higher even powers are not reducible to a single cosine in the smooth limit. For example, $\cos^4x=3/8+(1/2)\cos2x+(1/8)\cos4x$, which contains a positive dc component and multiple even harmonics. Thus, while $n=2$ maps to a conventional single-cosine modulation with a $\lambda$-dependent hopping shift, the quantitative phase boundaries for $n\geq4$ depend on the full harmonic content through the zero-mode logarithmic average.  Figs.~\ref{Fig3}(d)--\ref{Fig3}(f) show representative curves of $\nu$ and $Q$ versus $\lambda$. For $t_1=-2$, two transition points are clearly visible, confirming the emergence of a TAI-like reentrant topological phase. For $t_1=-0.5$, only one transition from the topologically nontrivial phase to the trivial phase is observed. By contrast, for $t_1=1.5$ the system remains trivial for all $\lambda$. Having established the phase diagrams in the smooth quasiperiodic limit, we now turn to the opposite extreme and examine how the topological structure changes when the modulation profile approaches its sign-function limit.

\subsection{$\beta \to \infty$ limit}\label{sec:betainf}

We next consider the opposite limit $\beta\to \infty$, for which the intracell hopping becomes $\omega_j=t_1+\lambda\,\mathrm{sgn}\!\left[\cos^n(2\pi\alpha j)\right]$. The inverse localization length of the zero-energy mode is then given by
\begin{equation}
\label{eq14}
\nu=\left|\lim_{N\to\infty}\frac{1}{N}\sum_{j=1}^{N}\ln\left|t_1+\lambda\,\mathrm{sgn}\!\left[\cos^n\!\left(2\pi\alpha(j+j_0)\right)\right]\right|\right|,
\end{equation}
where $j_0$ is an arbitrary initial site index, as in Eq.~(\ref{eq5}). Using the same argument as in Eq.~(\ref{eq6}), the inverse localization length can be expressed as
\begin{equation}
\label{eq15}
\nu=\left|\frac{1}{2\pi}\int_{0}^{2\pi}\ln\left|t_1+\lambda\,\mathrm{sgn}\!\left(\cos^n x\right)\right|dx\right|.
\end{equation}
As in the $\beta\to 0$ limit, we discuss the odd- and even-$n$ cases separately.

\begin{figure}
\centering
\includegraphics[width=0.7\linewidth]{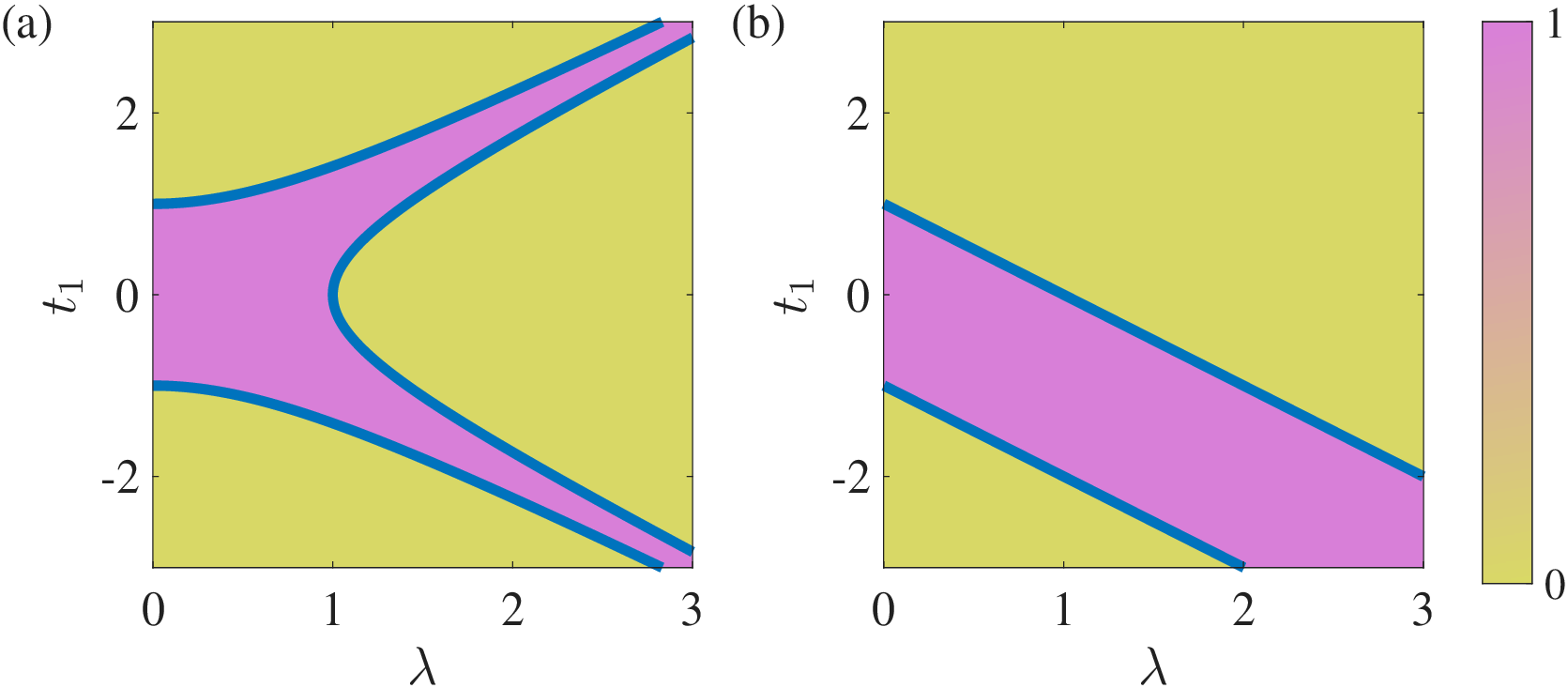}
\caption{\label{Fig4}Topological phase diagrams in the $\beta\to \infty$ limit. The colored regions are obtained from the numerical topological indicator $Q$, and the blue solid lines denote the analytical phase boundaries determined by $\nu=0$. (a) Odd $n$. (b) Even $n$. Here $N=610$ unit cells are used, corresponding to $L=1220$ lattice sites.}
\end{figure}

For odd $n$, one has $\mathrm{sgn}(\cos^n x)=\mathrm{sgn}(\cos x)$, so that the quasiperiodic modulation is reduced to a binary square-wave form with $\Delta_j=\pm\lambda$. Since the two values $\pm\lambda$ occur with equal weight in the thermodynamic limit, Eq.~(\ref{eq15}) yields the inverse localization length
\begin{equation}
\label{eq16}
\nu=\frac{1}{2}\left|\ln\left|t_1^2-\lambda^2\right|\right|.
\end{equation}
The corresponding phase boundaries are determined from $\nu=0$ as
\begin{equation}
\label{eq17}
|t_1^2-\lambda^2|=1.
\end{equation}
Fig.~\ref{Fig4}(a) shows the phase diagram in the $(t_1,\lambda)$ plane for the odd-$n$ case. The analytical boundaries from Eq.~(\ref{eq17}) are plotted as blue solid curves and agree well with the numerical results. In the interval $|t_1|>1$, the system exhibits a reentrant transition sequence: it is trivial at weak modulation, enters a TAI-like reentrant topological phase at intermediate $\lambda$, and returns to the trivial phase when $\lambda$ is further increased. Compared with the $\beta\to 0$ limit, the topological region in the present case extends to progressively larger $\lambda$ as $|t_1|$ increases, indicating that stronger modulation can still support a topological phase provided that the uniform hopping is tuned simultaneously.

For even $n$, one has $\mathrm{sgn}(\cos^n x)=1$, so that all even-$n$ cases reduce to a spatially uniform modulation with $\Delta_j=\lambda$. In this case Eq.~(\ref{eq15}) simplifies to
\begin{equation}
\label{eq18}
\nu=\left|\ln|t_1+\lambda|\right|,
\end{equation}
and the phase boundary is determined by
\begin{equation}
\label{eq19}
|t_1+\lambda|=1.
\end{equation}
It should be emphasized that, in the even-$n$ case, the $\beta\to\infty$ limit is qualitatively different from a genuine quasiperiodic or disorder-induced TAI mechanism, where the modulation becomes spatially uniform and simply shifts the intracell hopping from $t_1$ to $t_1+\lambda$. Therefore, the corresponding reentrant transition in this limiting case should be regarded as a uniform hopping-shift-induced transition rather than a genuine quasiperiodicity-induced topological Anderson mechanism. Nevertheless, this limit is useful as an endpoint of the continuous interpolation controlled by $\beta$, where the even-$n$ full modulation profile becomes uniform. Fig.~\ref{Fig4}(b) shows the corresponding phase diagram for the even-$n$ case. The analytical boundaries from Eq.~(\ref{eq19}) are plotted as blue solid lines and are in excellent agreement with the numerical phase boundaries. In the region $t_1<-1$, the system again undergoes a reentrant trivial--topological--trivial transition as $\lambda$ increases. In contrast to the odd-$n$ case, however, the topological region has a constant width, reflecting the fact that the $\beta\to\infty$ limit for even $n$ corresponds to an effectively uniform modulation rather than a binary aperiodic one.

The two limiting cases thus reveal a clear even--odd contrast in the topological phase structure. In the following, we examine whether this contrast persists throughout the continuous interpolation between the two limits.

\begin{figure}
\includegraphics[width=\linewidth]{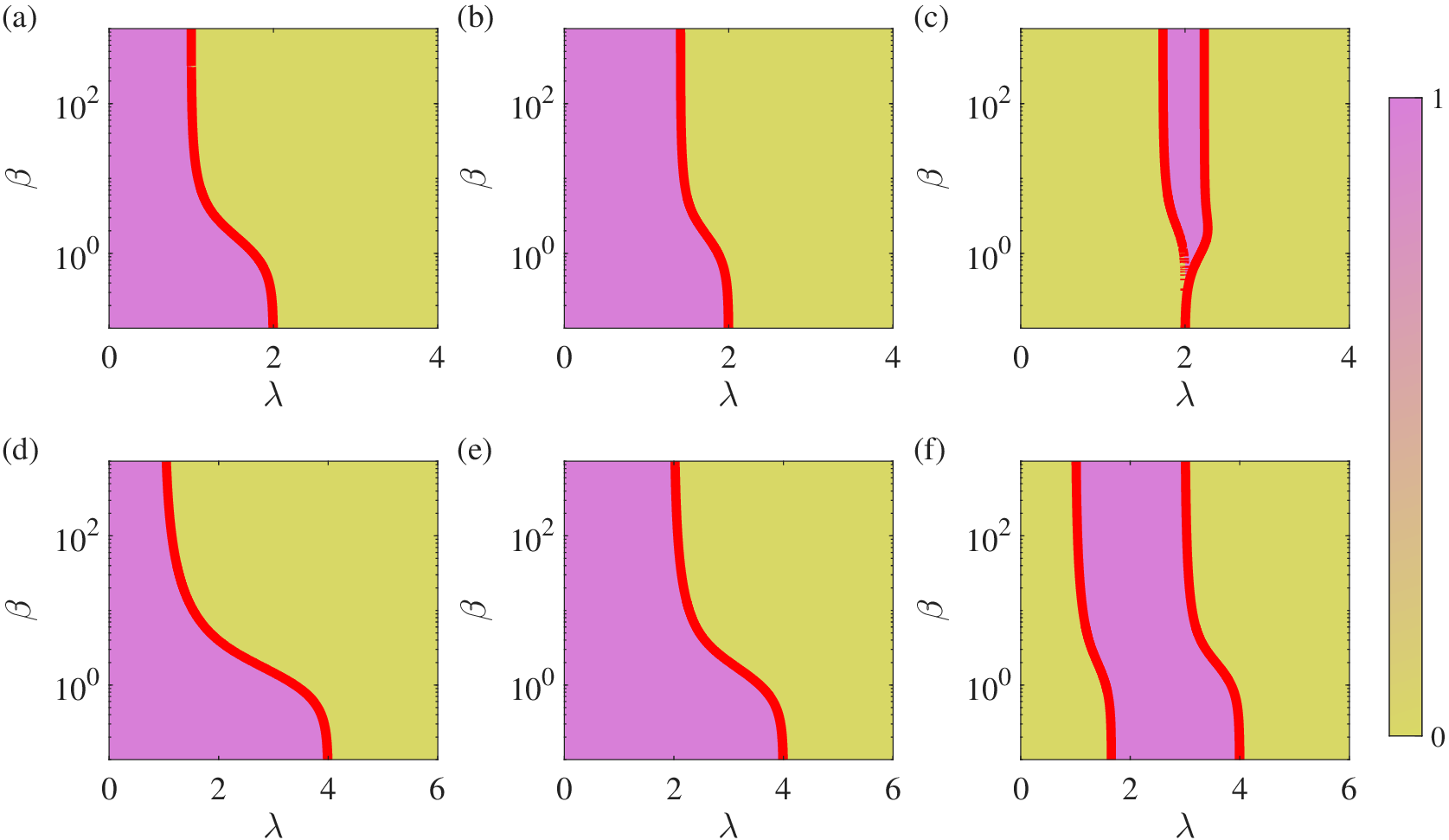}
\caption{\label{Fig5}Topological phase diagrams in the $(\beta,\lambda)$ plane for continuously varying $\beta$. The colored regions are obtained from the numerical topological indicator $Q$, and the red solid lines denote the phase boundaries determined from the zero-mode condition $\nu(\beta)=0$. (a)--(c) $n=1$ with $t_1=0$, 1, and 2, respectively. (d)--(f) $n=2$ with $t_1=0$, $-1$, and $-2$, respectively. Here $N=610$ unit cells are used, corresponding to $L=1220$ lattice sites.}
\end{figure}

\subsection{Continuously varying $\beta$}\label{sec:beta_cont}

We now turn to the intermediate regime of finite $\beta$, which continuously interpolates between the two analytically tractable limits discussed above. The central question is whether the even--odd dependence of the topological phase structure identified in those limits remains robust throughout this interpolation.

For arbitrary positive integer $n$, the intracell hopping at finite $\beta$ takes the form
\begin{equation}
\label{eq25}
\omega_j=t_1+\lambda \frac{\tanh\!\left[\beta \cos^n(2\pi\alpha j)\right]}{\tanh\beta}.
\end{equation}
Using the same equidistribution argument as in Eq.~(\ref{eq6}), the inverse localization length of the zero-energy mode is
\begin{equation}
\label{eq26}
\nu(\beta)=\left|\frac{1}{2\pi}\int_{0}^{2\pi}\ln\left|t_1+\lambda\frac{\tanh(\beta\cos^n x)}{\tanh\beta}
\right|dx\right|.
\end{equation}
The topological phase boundary is determined implicitly by the zero-mode delocalization condition
\begin{equation}
\label{eq27}
\nu(\beta)=0.
\end{equation}
Although this equation does not generally admit a closed-form solution for finite $\beta$, it can be evaluated numerically with high precision and correctly reproduces the analytical phase boundaries in both limiting cases. The topological character of the regions separated by these boundaries is determined from the real-space indicator $Q$.

Fig.~\ref{Fig5} summarizes the resulting phase diagrams in the $(\beta,\lambda)$ plane, with $n=1$ and $n=2$ taken as representative odd- and even-$n$ cases, respectively. For odd $n$, as shown in  Figs.~\ref{Fig5}(a)--\ref{Fig5}(c), the three panels correspond to qualitatively distinct clean-limit regimes: $t_1=0$ lies in the clean topological phase, $t_1=1$ is the clean critical point, and $t_1=2$ lies in the clean trivial phase. Only for $t_1=2$, which starts from the clean trivial regime $|t_1|>1$, does a reentrant TAI-like topological window emerge at intermediate modulation strength. For even $n$, as shown in  Figs.~\ref{Fig5}(d)--\ref{Fig5}(f), the panels correspond to $t_1=0$ (clean topological), $t_1=-1$ (clean critical), and $t_1=-2$ (clean trivial on the negative side). A reentrant TAI-like region appears only for $t_1=-2$, consistent with the even-$n$ mechanism.

These results confirm that the reentrant topological phase persists throughout the continuous variation of $\beta$ whenever the corresponding modulation window exists. While the precise location and width of the reentrant window evolve smoothly with $\beta$, the underlying transition pattern remains unchanged. For positive $\lambda$, odd-$n$ modulations can induce reentrance from the clean trivial regime $|t_1|>1$, whereas even-$n$ modulations do so only from the negative clean trivial side $t_1<-1$. Thus, varying $\beta$ continuously deforms the phase boundary determined by $\nu=0$ without changing the even--odd structure of the phase diagrams.

\subsection{Even--odd effect: dc component and support of the full modulation profile}\label{sec:parity}

The results of Sections \ref{sec:beta0}--\ref{sec:beta_cont} show that the even--odd dependence of the phase diagrams is controlled by the full modulation profile
$f_n(x;\beta)=\tanh\!\left(\beta\cos^n x\right)/\tanh\beta$. For odd $n$, $f_n(x+\pi;\beta)=-f_n(x;\beta)$, so this profile has zero spatial average and changes sign about zero. For positive $\lambda$, the corresponding modulation therefore provides both positive and negative corrections to the bare hopping, allowing reentrant topology to be induced from the clean trivial regime corrections to the bare hopping, allowing reentrant topology to be induced from the clean trivial regime $|t_1|>1$.

For even $n$, $f_n(x;\beta)\geq0$ and the spatial average is positive. After subtracting this average, the remaining fluctuation can change sign, as seen explicitly in the $n=2$ smooth-limit decomposition discussed above. However, the full modulation profile remains non-negative. The relevant even--odd distinction is therefore not the presence or absence of sign-changing fluctuations, but whether the full profile is sign-changing or non-negative. The positive average acts as a $\lambda$-dependent shift of the effective intracell hopping, which can compensate a negative trivial hopping $t_1<-1$ but cannot reduce a positive trivial hopping $t_1>1$. This explains why the even-$n$ reentrant window for positive $\lambda$ is confined to the negative-$t_1$ side.

The $n=2$, $\beta\to0$ case is a special single-cosine mapping along a constrained path in the effective parameter space, whereas higher even powers already contain additional harmonics in the smooth limit and their quantitative phase boundaries depend on the full waveform through the zero-mode logarithmic average. Varying $\beta$ changes the detailed location and width of the reentrant windows, but it does not change the basic even--odd mechanism set by the dc component and support of the full modulation profile.

\subsection{Relation to bulk localization properties}\label{sec:bulk_main}

Previous studies have shown that quasiperiodic SSH chains may host topological Anderson-like phases with extended, intermediate, or localized bulk states, and that mobility-edge-like structures can coexist with nontrivial
topology~\cite{FrontiersZhanpeng2025,Ann2022Lu,PRA2022Tang}. To clarify the situation in the present model, we analyze the inverse participation ratios, fractal dimensions, and representative wave-function profiles in Appendix 
\ref{sec:bulk}. The results show that, in the smooth integer-power limit $\beta\to0$, the bulk states can display extended, localized, and intermediate localization regimes, including mobility-edge-like behavior and
regions containing fractal-like states. In particular, changes in the bulk localization properties show visible correlations with parts of the topological phase boundaries. By contrast, in the sign-function limit
$\beta\to\infty$, such correlations become much weaker. These results indicate that mobility-edge-like and fractal-like bulk structures may coexist with the reentrant topological windows. However, the topological phase
boundaries discussed in the main text are governed by the zero-energy transfer criterion $\nu=0$ and do not require a simultaneous localization transition of all bulk states.

\begin{figure}
\includegraphics[width=\linewidth]{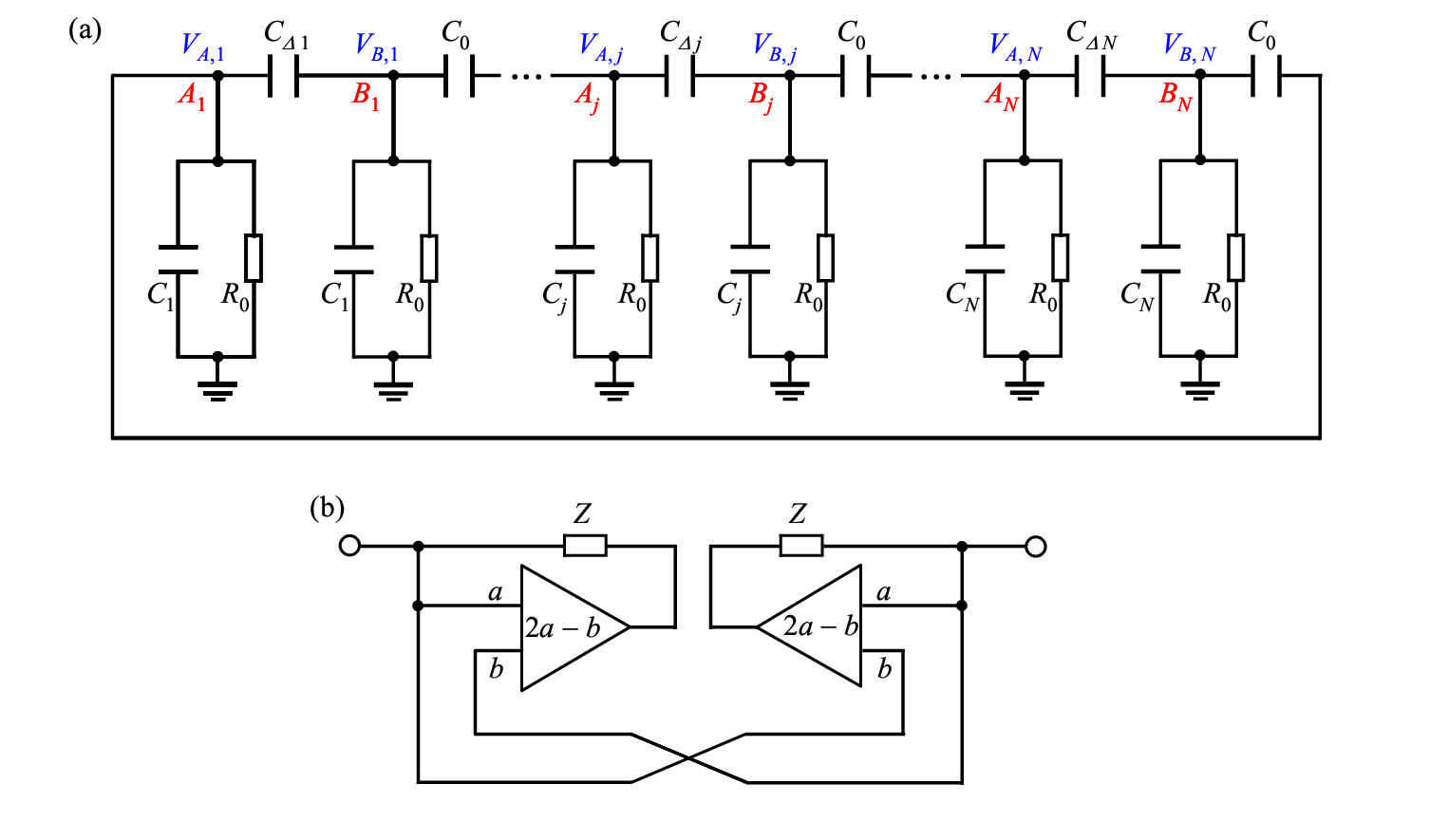}
\caption{\label{Fig6}Electrical-circuit implementation of the model. (a) Main circuit under periodic boundary conditions. Adjacent nodes represent the $A$ and $B$ sublattices of the SSH chain. (b) Two-terminal active circuit used to realize an effective negative capacitance. Here $Z$ denotes an arbitrary impedance element.}
\end{figure}

\section{Electrical-circuit implementation}\label{sec:experiment}

The reentrant topological phase predicted above can be implemented and detected in an electrical-circuit platform. To this end, we propose the circuit shown in Fig.~ \ref{Fig6}, which realizes the Hamiltonian in Eq.~(\ref{eq1}). As illustrated in Fig.~\ref{Fig6}(a), adjacent nodes in the AC circuit chain represent the $A$ and $B$ sublattices of the SSH model. The capacitors $C_{\Delta j}$ and $C_0$ simulate the intracell and intercell hoppings, respectively. Here $C_{\Delta j}$ corresponds to the full intracell hopping amplitude $\omega_j=t_1+\Delta_j$, while $C_0$ represents the uniform intercell hopping. The node voltages $V_{A,j}$ and $V_{B,j}$ correspond to the wave-function amplitudes on the two sublattices. In addition, the grounded capacitor $C_j$ is introduced to balance the on-site terms, while the resistor $R_0$ is included to protect the circuit. In the present implementation, $C_0$ serves as the reference capacitance corresponding to the intercell hopping scale.

When an effective negative capacitance is required for $C_{\Delta j}$ or $C_j$, it can be realized by the two-terminal active circuit shown in Fig.~\ref{Fig6}(b), which consists of two capacitors and two operational
amplifiers~\cite{njp2024Liu}. For an AC driving frequency $\Omega$, the Hamiltonian in Eq.~(\ref{eq1}) is mapped onto the circuit Laplacian $\mathcal{J}(\Omega)$, which relates the input-current vector $I(\Omega)$ to the
node-voltage vector $V(\Omega)$ through
\begin{equation}
\label{eq28}
I(\Omega)=\mathcal{J}(\Omega)V(\Omega).
\end{equation}

Applying Kirchhoff's current law to the bulk and boundary nodes yields
\begin{equation}
\label{eq29}
\begin{aligned}
I_{A,j}={}&i\Omega C_{0}V_{B,j-1}+i\Omega C_{\Delta j}V_{B,j} -i\Omega \left( C_{0}+C_{\Delta j}+\frac{1}{i\Omega R_{0}}+C_{j} \right)V_{A,j}, \\
I_{B,j}={}&i\Omega C_{\Delta j}V_{A,j}+i\Omega C_{0}V_{A,j+1}-i\Omega \left( C_{0}+C_{\Delta j}+\frac{1}{i\Omega R_{0}}+C_{j} \right)V_{B,j},
\end{aligned}
\end{equation}
while the boundary cells under periodic boundary conditions obey
\begin{equation}
\label{eq31}
\begin{aligned}
I_{A,1}={}&i\Omega C_{0}V_{B,N}+i\Omega C_{\Delta 1}V_{B,1}-i\Omega \left( C_{0}+C_{\Delta 1}+\frac{1}{i\Omega R_{0}}+C_{1} \right)V_{A,1}, \\
I_{B,N}={}&i\Omega C_{\Delta N}V_{A,N}+i\Omega C_{0}V_{A,1} -i\Omega \left( C_{0}+C_{\Delta N}+\frac{1}{i\Omega R_{0}}+C_{N} \right)V_{B,N}.
\end{aligned}
\end{equation}
Here $I_{\sigma,j}$ and $V_{\sigma,j}$ denote the current and voltage at sublattice $\sigma\in\{A,B\}$ in the $j$th unit cell, respectively. By choosing $C_{\Delta j}=\omega_j C_0=(t_1+\Delta_j)C_0$, and $C_j=-(C_0+C_{\Delta j})$, the circuit Laplacian in the basis $(V_{A,1},V_{B,1},V_{A,2},V_{B,2},\ldots,V_{A,N},V_{B,N})^{T}$ takes the form
\begin{equation}
\label{eq33}
\mathcal{J}(\Omega)=i\Omega\mathcal{M}-\frac{1}{R_0}\mathcal{I},
\end{equation}
where $\mathcal{I}$ is the $2N\times 2N$ identity matrix, and the nonzero elements of $\mathcal{M}$ are given by
\begin{equation*}
\mathcal{M}_{2j-1,2j}=\mathcal{M}_{2j,2j-1}=C_{\Delta j}\qquad (j=1,\dots,N),
\end{equation*}
\begin{equation*}
\mathcal{M}_{2j,2j+1}=\mathcal{M}_{2j+1,2j}=C_0\qquad (j=1,\dots,N-1),
\end{equation*}
together with the periodic boundary terms $\mathcal{M}_{2N,1}=\mathcal{M}_{1,2N}=C_0$, while all remaining matrix elements vanish. The circuit Laplacian therefore reproduces the target SSH Hamiltonian up to the overall prefactor $i\Omega C_0$ and the uniform diagonal shift $-R_0^{-1}\mathcal{I}$.

While Fig.~\ref{Fig6}(a) illustrates the periodic-boundary implementation used to reproduce the bulk Hamiltonian, the detection of topological boundary modes can be carried out in the corresponding open-boundary circuit
obtained by removing the end-to-end connection. In this case, the spatial distribution of the eigenstates can be probed directly through voltage measurements at the circuit nodes~\cite{njp2024Liu}.

To experimentally detect the reentrant topological phase, one may measure the node-voltage distribution and the two-point impedance under open boundary conditions. In the topologically nontrivial regime, the circuit
Laplacian supports boundary-localized eigenvectors corresponding to the SSH edge modes. When an AC current is injected near the boundary, the voltage response is therefore strongly concentrated near the edge nodes in the
low-loss regime. By contrast, such boundary-localized voltage profiles are absent in the trivial regime. A complementary probe is provided by the two-point impedance between boundary nodes. The impedance matrix is
determined by the inverse of the circuit Laplacian and is therefore sensitive to eigenmodes with small admittance. In the topological regime, the boundary-localized mode produces an enhanced edge response compared with the
trivial regime. Thus, by varying the modulation strength $\lambda$, the appearance and disappearance of an enhanced boundary response can serve as an experimental signature of the reentrant trivial--topological--trivial
transition predicted in this model. In addition, one may inject an AC current from one boundary node and measure the voltage response at the opposite boundary, thereby obtaining a two-terminal probe of the reentrant
topological phase. These measurements are well within the capabilities of existing topolectrical-circuit platforms~\cite{njp2024Liu}. Moreover, practical experiments require the system to be robust against manufacturing
errors of the circuit components, which we discuss in Appendix \ref{sec:robustness}.

\section{Conclusion}\label{sec:conclusion}

In conclusion, we have studied a one-dimensional SSH chain with integer-power quasiperiodic intracell hopping modulation. By combining the zero-mode inverse-localization-length criterion $\nu=0$ with the real-space topological indicator $Q$, we obtained consistent analytical and numerical phase diagrams in the smooth quasiperiodic limit, the sign-function limit, and the finite-$\beta$ interpolation between them. These results show that deterministic quasiperiodic hopping modulation can induce topological Anderson-like reentrant phases within finite parameter windows.

The main finding is an even--odd classification controlled by the dc component and support of the full integer-power modulation profile, rather than by spatial inversion symmetry. For positive modulation strength, odd powers generate a zero-mean sign-changing profile and can induce reentrant topology from the clean trivial regime $|t_1|>1$. Even powers generate a positive-mean non-negative profile, which can compensate a negative clean trivial hopping $t_1<-1$ but cannot reduce a positive trivial hopping $t_1>1$. Although sign-changing fluctuations may appear after subtracting the positive average, the full even-power modulation profile remains non-negative. Thus, the even--odd classification should be understood in terms of the full modulation profile and, in the even-power case, the corresponding constrained path in effective parameter space.

We also showed that mobility-edge-like and fractal-like bulk structures may coexist with the reentrant topological windows, although the topological phase boundaries are governed by the zero-energy transfer criterion and do not require a simultaneous localization transition of all bulk states. Finally, the phase diagrams are robust against moderate random hopping fluctuations, and the proposed electrical-circuit platform provides a feasible route to observe the reentrant trivial--topological--trivial transition through boundary voltage and impedance measurements.

\printcredits

\section*{Declaration of competing interest}
The authors declare that they have no known competing financial interests or personal relationships that could have appeared to influence the work reported in this paper.

\section*{Acknowledgments}
Z. X. is supported by Quantum Science and Technology-National Science and Technology Major Project (Grant No. 2025ZD0300400), the National Natural Science Foundation of China (Grant Nos. 12375016 and 12461160324), and Beijing National Laboratory for Condensed Matter Physics (No. 2023BNLCMPKF001). Y. N. is supported by the National Training Program of Innovation for Undergraduates under Grant No. 202510108034.




\section*{Appendices}

\appendix{}
\section[pfx={Appendix\space}]{Calculation of the zero-mode inverse localization length}\label{sec:calculation}

The following subsections present exact analytical expressions for the zero-mode inverse localization length in the $\beta\to 0$ limit for $n=3$, $n=2$, and $n=4$, respectively. The derivation for $n=4$ (Appendix \ref{appendix3}) makes use of the intermediate result established in  Appendix \ref{appendix2}.

\subsection{Zero-mode inverse localization length for the $\beta\to 0$ limit with $n=3$}\label{appendix1}

For $\beta\to 0$ and $n=3$, the inverse localization length of the zero-energy mode is
\numberwithin{equation}{section}
\setcounter{equation}{0}
\begin{equation}
\nu=
\left|
\frac{1}{2\pi}\int_{0}^{2\pi}\ln\left|t_1+\lambda \cos^3 x\right|dx
\right|.
\end{equation}
To evaluate this integral, we introduce the complex variable $z=e^{ix}$, for which
\begin{equation}
\cos x=\frac{1}{2}\left(z+\frac{1}{z}\right),
\qquad
dx=\frac{dz}{iz}.
\end{equation}
Using
\begin{equation}
\cos^3x=\frac{1}{8}\left(z+\frac{1}{z}\right)^3,
\end{equation}
we rewrite the integrand as
\begin{equation}
t_1+\lambda\cos^3x
=
\frac{1}{z^3}
\left[
\frac{\lambda}{8}(z^2+1)^3+t_1 z^3
\right].
\end{equation}
We therefore define the polynomial
\begin{equation}
P(z)=\frac{\lambda}{8}(z^2+1)^3+t_1 z^3.
\end{equation}
Since $|z|=1$ on the integration contour, Jensen's formula gives
\begin{equation}
\frac{1}{2\pi}\int_{0}^{2\pi}\ln\left|t_1+\lambda\cos^3x\right|dx
=
\ln\left|\frac{\lambda}{8}\right|
+
\sum_{|z_j|>1}\ln|z_j|,
\end{equation}
where $z_j$ are the six roots of $P(z)=0$.

Next, we use the palindromic structure of $P(z)$. Dividing by $z^3$, one obtains
\begin{equation}
\frac{P(z)}{z^3}
=
\frac{\lambda}{8}\left(z+\frac{1}{z}\right)^3+t_1.
\end{equation}
Introducing
\begin{equation}
w=z+\frac{1}{z},
\end{equation}
the equation $P(z)=0$ reduces to
\begin{equation}
\frac{\lambda}{8}w^3+t_1=0,
\qquad\text{or}\qquad
w^3=-\frac{8t_1}{\lambda}.
\end{equation}
Let
\begin{equation}
s=\left(\frac{t_1}{\lambda}\right)^{1/3},
\end{equation}
where the real cube root is taken for real $t_1/\lambda$. The three roots of the cubic equation can then be written as
\begin{equation}
w_0=-2s,
\qquad
w_1=s(1+i\sqrt{3}),
\qquad
w_2=s(1-i\sqrt{3}).
\end{equation}

For each $w_\mu$, the corresponding $z$ roots satisfy
\begin{equation}
z^2-w_\mu z+1=0.
\end{equation}
Because the two roots always obey $z_+z_-=1$, one root lies outside the unit circle and the other lies inside it, except at the transition point where they sit on the unit circle. Therefore, each $w_\mu$ contributes the logarithm of the modulus of the root with $|z|>1$.

We first consider the real root $w_0=-2s$. The corresponding quadratic equation is
\begin{equation}
z^2+2sz+1=0.
\end{equation}
For $|s|<1$, both roots lie on the unit circle and hence give no contribution to the inverse localization length. For $|s|>1$, the root outside the unit circle has modulus
\begin{equation}
|z|=|s|+\sqrt{s^2-1},
\end{equation}
so that its contribution is
\begin{equation}
\delta_0=
\left\{
\begin{aligned}
&\ln\left(|s|+\sqrt{s^2-1}\right), && |s|>1,\\
&0, && |s|<1.
\end{aligned}
\right.
\end{equation}

We next consider the complex-conjugate pair $w_1$ and $w_2$. For each of them, the two roots of
\begin{equation}
z^2-w_\mu z+1=0
\end{equation}
satisfy $z_+z_-=1$, so that the contribution to Jensen's formula is given by the logarithm of the modulus of the root with $|z|>1$. For the pair $w_1$ and $w_2$, these contributions are equal and can be written compactly as
\begin{equation}
\delta_1+\delta_2
=
2\ln\left|
\frac{w_1+\sqrt{w_1^2-4}}{2}
\right|,
\end{equation}
where the branch of the square root is chosen continuously such that
\begin{equation}
\left|
\frac{w_1+\sqrt{w_1^2-4}}{2}
\right|>1.
\end{equation}
Since $w_2=\overline{w_1}$, the total contribution is manifestly real.

Collecting all contributions, we obtain the exact zero-mode inverse localization length in the form
\begin{equation}
\nu=\left|\ln\left|\frac{\lambda}{8}\right|+2\ln\left|\frac{w_1+\sqrt{w_1^2-4}}{2}\right|+\delta_0\right|,
\end{equation}
with
\begin{equation}
w_1=s(1+i\sqrt{3}),
\qquad
s=\left(\frac{t_1}{\lambda}\right)^{1/3}.
\end{equation}
Equivalently, the result can be written explicitly as
\begin{equation}
\nu=
\left\{
\begin{aligned}
&
\left|
\ln\left|\frac{\lambda}{8}\right|
+
2\ln\left|
\frac{w_1+\sqrt{w_1^2-4}}{2}
\right|
+
\ln\left(|s|+\sqrt{s^2-1}\right)
\right|,
&& |t_1/\lambda|>1,\\
&
\left|
\ln\left|\frac{\lambda}{8}\right|
+
2\ln\left|
\frac{w_1+\sqrt{w_1^2-4}}{2}
\right|
\right|,
&& 0<|t_1/\lambda|<1,
\end{aligned}
\right.
\end{equation}
where the boundary case $|t_1/\lambda|=1$ (i.e.,~$|s|=1$) follows by continuity from either branch, since  $\delta_0=\ln(|s|+\sqrt{s^2-1})\to 0$ as $|s|\to 1^+$. This representation is sufficient for determining the phase boundary from the condition $\nu=0$ and can be evaluated numerically without ambiguity for all real $t_1/\lambda$.

As a consistency check, at the special point $t_1=0$ one has $s=0$ and hence $w_1=0$. The above expression reduces to
\begin{equation}
\nu=\left|\ln\left|\frac{\lambda}{8}\right|\right|,
\end{equation}
which agrees with the direct evaluation
\begin{equation}
\frac{1}{2\pi}\int_0^{2\pi}\ln|\lambda\cos^3x|\,dx
=
\ln|\lambda|-3\ln2
=
\ln\left|\frac{\lambda}{8}\right|.
\end{equation}
Therefore, the prefactor is fixed exactly by the analysis itself, and no additional fitting constant is needed.

\subsection{Zero-mode inverse localization length for the $\beta\to 0$ limit with $n=2$}\label{appendix2}

For $\beta\to 0$ and $n=2$, the inverse localization length of the 
zero-energy mode is
\begin{equation}
\nu=\left|\frac{1}{2\pi}\int_{0}^{2\pi}\ln\left|t_1+\lambda \cos^2 x
\right|dx\right|.
\end{equation}
For $t_1\neq 0$, the logarithmic integral can be written as
\begin{equation}
\frac{1}{2\pi}\int_{0}^{2\pi}\ln\left|t_1+\lambda \cos^2 x\right|dx
=\ln|t_1|+h\!\left(\frac{\lambda}{t_1}\right),
\end{equation}
where
\begin{equation}
h(a)=\frac{1}{2\pi}\int_{0}^{2\pi}\ln\left|1+a\cos^2 x\right|dx,\qquad a=\frac{\lambda}{t_1}.
\end{equation}
Using $\cos^2 x=(1+\cos 2x)/2$, one obtains
\begin{equation}
1+a\cos^2x=A+B\cos 2x,\qquad A=1+\frac{a}{2},\qquad B=\frac{a}{2}.
\end{equation}
Since the average over $2x$ is equivalent to the average over one full 
period, $h(a)$ reduces to the standard integral~\cite{PRB2019Longhi,PRB2021Longhi}
\begin{equation}
\frac{1}{2\pi}\int_{0}^{2\pi}\ln|A+B\cos\theta|\,d\theta
=\begin{cases}
\ln\!\left|\dfrac{A+\sqrt{A^2-B^2}}{2}\right|, & A>|B|,\\[3pt]
\ln\!\left|\dfrac{B}{2}\right|, & |A|<|B|.
\end{cases}
\end{equation}
Since $A^2-B^2=1+a$, the present parametrization gives $A>|B|$ for 
$a>-1$, whereas $|A|<|B|$ for $a<-1$. Therefore, for $a>-1$ one obtains
\begin{equation}
h(a)=\ln\!\left|\frac{1+a/2+\sqrt{1+a}}{2}\right|
=2\ln|1+\sqrt{1+a}|-\ln 4.
\end{equation}
For $a<-1$, one obtains
\begin{equation}
h(a)=\ln\!\left|\frac{a}{4}\right|.
\end{equation}
The two expressions coincide by continuity at $a=-1$.
Therefore, the zero-mode inverse localization length is
\begin{equation}
\nu=
\left\{
\begin{aligned}
&\left|2\ln\left|1+\sqrt{\frac{\lambda}{t_1}+1}\right|
+\ln|t_1|-\ln 4\right|,
&& \lambda/t_1>-1,\\
&\left|\ln\left|\frac{\lambda}{4}\right|\right|,
&& \lambda/t_1<-1.
\end{aligned}
\right.
\end{equation}
For $t_1=0$, the result follows directly from
\begin{equation}
\frac{1}{2\pi}\int_0^{2\pi}\ln|\lambda\cos^2x|\,dx
=\ln|\lambda|-2\ln 2=\ln\!\left|\frac{\lambda}{4}\right|.
\end{equation}

\subsection{Zero-mode inverse localization length for the $\beta\to 0$ limit with $n=4$}\label{appendix3}

For $\beta\to 0$ and $n=4$, the inverse localization length of the zero-energy mode is
\begin{equation}
\nu=\left|\frac{1}{2\pi}\int_{0}^{2\pi}\ln\left|t_1+\lambda \cos^4 x\right|dx\right|.
\end{equation}
We evaluate the integral in different parameter regions. For compactness, we define
\begin{equation*}
\kappa=\sqrt{-\lambda/t_1}, 
\end{equation*}
where the principal branch of the square root is used when $\kappa$ is complex.

For $\lambda/t_1<-1$, one has $\kappa>1$, and the integrand can be factorized as
\begin{equation}
1+\frac{\lambda}{t_1}\cos^4 x
=
\left(1+\kappa\cos^2 x\right)\left(1-\kappa\cos^2 x\right).
\end{equation}
The first factor $1+\kappa\cos^2 x$ corresponds to the parameter value $a=\kappa>0$, which falls in the $a>-1$ branch of Appendix \ref{appendix2}. The second factor $1-\kappa\cos^2 x$ corresponds to $a=-\kappa<-1$ (since $\kappa>1$), which falls in the $a<-1$ branch. Applying the respective results from Appendix \ref{appendix2} to each factor and combining the two contributions, the zero-mode inverse localization length becomes
\begin{equation}
\begin{aligned}
\nu
&=\left|\frac{1}{2\pi}\int_{0}^{2\pi}\ln\left|1+\kappa\cos^2 x\right|dx
+\frac{1}{2\pi}\int_{0}^{2\pi}\ln\left|1-\kappa\cos^2 x\right|dx
+\ln|t_1| \right|\\
&=\left|\bigl[2\ln|1+\sqrt{1+\kappa}|-\ln 4\bigr]
+\bigl[\ln\kappa-\ln 4\bigr]
+\ln|t_1| \right|\\
&=\left|2\ln|1+\sqrt{1+\kappa}|+\ln|\kappa|+\ln|t_1|-\ln 16\right|.
\end{aligned}
\end{equation}

For $\lambda/t_1>-1$, the zero-mode inverse localization length takes the form
\begin{equation}
\nu=
\left|2\ln|1+\sqrt{1+\kappa}|
+2\ln|1+\sqrt{1-\kappa}|
+\ln|t_1|-\ln 16\right|,
\end{equation}
where $\kappa=(-\lambda/t_1)^{1/2}$. For $\lambda/t_1<0$, $\kappa$ is real and positive. For $\lambda/t_1>0$, $\kappa=i|\kappa|$ is purely imaginary, so that $1+\sqrt{1-\kappa}$ and $1+\sqrt{1+\kappa}$ become complex; in this case the modulus $|\cdot|$ is taken before the logarithm, which ensures that $\nu$ remains real and non-negative for all parameter values. Throughout this appendix, the principal branch of the square root is used; the final expressions are understood through their absolute values, which makes the zero-mode inverse localization length single-valued and real.

Therefore, the zero-mode inverse localization length can be summarized as
\begin{equation}
\nu=
\left\{
\begin{aligned}
&\left|
2\ln|1+\sqrt{1+\kappa}|
+2\ln|1+\sqrt{1-\kappa}|
+\ln|t_1|-\ln 16
\right|,
&& \lambda/t_1>-1,\\
&\left|
2\ln|1+\sqrt{1+\kappa}|
+\ln|\kappa|
+\ln|t_1|-\ln 16
\right|,
&& \lambda/t_1<-1,
\end{aligned}
\right.
\end{equation}
where in both cases $\kappa=(-\lambda/t_1)^{1/2}$, with the principal branch of the square root used throughout. The modulus $|\cdot|$ is taken before each logarithm to ensure that $\nu$ remains real and non-negative for all parameter values.

\section{Bulk localization properties}\label{sec:bulk}

In some reciprocal or nonreciprocal models with cosine-type quasiperiodic modulation, the energy spectrum exhibits rich structures~\cite{Ann2022Lu,PRA2022Tang,PRResearch2024Li,PRL2022majie}, such as mobility edges and
multifractal states. In the $\beta\to0$ limit, our modulation reduces to a smooth integer-power quasiperiodic profile. We therefore analyze the bulk localization properties in this limit and compare them with those in the
$\beta\to\infty$ limit and at finite $\beta$. The discussion below uses $n=2$ and $n=3$ as representative examples. We characterize the localization properties using the real- and momentum-space inverse
participation ratios (IPRs), the diagnostic quantity $\eta$, and the fractal dimension $D_2$.

Following Refs.~\cite{Ann2022Lu,PRA2022Tang}, for the $l$th normalized eigenstate, the IPR in representation $\mu$ is defined as
\begin{equation}
I_{\mu}^{(l)}=\sum_{j=1}^{L}\left|\phi_{\mu}^{(l)}(j)\right|^4,\qquad \mu=x,k ,
\end{equation}
where $\mu=x$ and $\mu=k$ denote the real-space and momentum-space representations, respectively. The momentum-space amplitude $\phi_k^{(l)}(j)$ is obtained from the real-space wave function by a discrete Fourier
transform. We also use the averaged IPR~\cite{PRL2013Cai},
\begin{equation}
\left\langle I_{\mu}\right\rangle=\frac{1}{L}\sum_{l=1}^{L}I_{\mu}^{(l)} .
\end{equation}
For an extended regime, $\langle I_x\rangle\sim L^{-1}$ and $\langle I_k\rangle\sim O(1)$, whereas for a localized regime, $\langle I_x\rangle\sim O(1)$ and $\langle I_k\rangle\sim L^{-1}$.

To identify possible intermediate regimes, we further introduce the dimensionless quantity~\cite{PRA2022Tang}
\begin{equation}
\eta=\log_{10}\left(\langle I_x\rangle\langle I_k\rangle\right).
\end{equation}
Regimes dominated by either localized or extended states usually exhibit smaller $\eta$ values than mixed or intermediate regimes. In addition, we compute the fractal dimension
\begin{equation}
D_2^{(l)}=-\frac{\ln I_x^{(l)}}{\ln L}.
\end{equation}
In the thermodynamic limit, $D_2^{(l)}\to0$ for localized states and $D_2^{(l)}\to1$ for extended states, while stable intermediate values indicate critical or multifractal behavior.

\begin{figure}
\includegraphics[width=\linewidth]{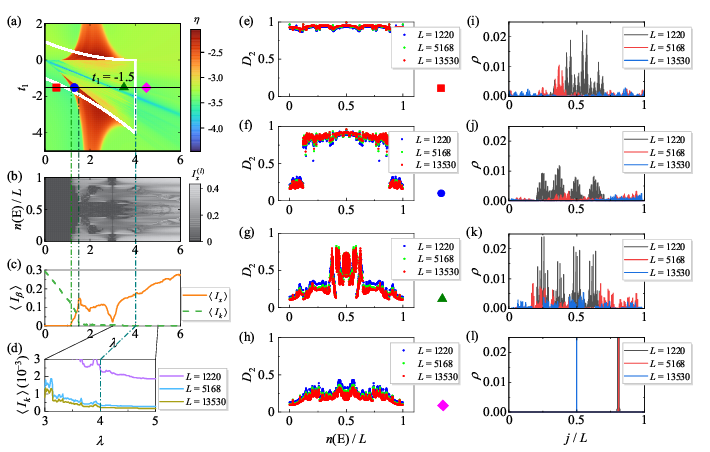}
\caption{\label{Fig7}Bulk localization properties in the $\beta\to0$ limit for $n=2$. When boundary conditions are required, PBC are used. (a) Color map of the diagnostic quantity $\eta$ in the $(t_1,\lambda)$ plane. The white solid curves denote the analytical topological phase boundaries, and the black solid line marks the cut $t_1=-1.5$. The colored symbols mark four representative points: extended regime (red square), intermediate regime with mobility-edge-like behavior (blue circle), intermediate regime with localized and fractal-like states (green triangle), and localized regime (magenta diamond). (b) Real-space IPR $I_x^{(l)}$ along the cut $t_1=-1.5$. (c) Averaged IPRs $\langle I_x\rangle$ and $\langle I_k\rangle$ as functions of $\lambda$. (d) Finite-size behavior of $\langle I_k\rangle$ near $\lambda=4$. (e)--(h) Fractal dimensions $D_2^{(l)}$ for the four representative points. (i)--(l) Probability-density distributions of the $0.3746L$th, $0.4141L$th, $0.4972L$th, and $0.09017L$th eigenstates selected from the parameter point in (g), shown for $L=1220,5168,13530$. The corresponding numbers of unit cells are $N=610,2584,6765$, respectively.}
\end{figure}

\subsection{Bulk localization properties for $n=2$ in the $\beta\to0$ limit}

Fig.~\ref{Fig7}(a) shows a color map of $\eta$ in the $(\lambda,t_1)$ plane for $n=2$ in the $\beta\to0$ limit. The enhanced-$\eta$ regions indicate mixed or intermediate localization behavior, while the white curves denote the analytical topological phase boundaries. The endpoints of the enhanced-$\eta$ regions are visibly correlated with the vertices of the topological phase boundaries, although they do not coincide exactly.

To examine the localization properties in more detail, we take the representative cut $t_1=-1.5$, as marked by the black line in Fig.~\ref{Fig7}(a). Along this cut, Figs.~\ref{Fig7}(b) and \ref{Fig7}(c) show that the system evolves with increasing $\lambda$ from an extended regime to an intermediate regime with mobility-edge-like behavior, then to a mixed regime containing localized and fractal-like states, and finally to a localized regime. The finite-size behavior of $\langle I_k\rangle$ near $\lambda=4$, shown in Fig.~\ref{Fig7}(d), further suggests that the bulk states on the two sides of this point have different localization characteristics. We also note that $\lambda=4$ coincides with one branch of the topological phase boundary for this cut.

Figs.~~\ref{Fig7}(e)--\ref{Fig7}(h) show the fractal dimensions $D_2^{(l)}$ for four representative parameter points. In Fig.~\ref{Fig7}(e), $D_2^{(l)}$ approaches unity with increasing system size, indicating an extended regime. In Fig.~\ref{Fig7}(h), $D_2^{(l)}$ approaches zero, indicating a localized regime. In Fig.~\ref{Fig7}(f), states in the middle part of the spectrum tend toward extended behavior, while those near the spectral edges tend toward localized behavior, supporting mobility-edge-like separation. In Fig.~\ref{Fig7}(g), the middle part of the spectrum exhibits stable intermediate $D_2$ values, while other states tend toward localized behavior. The representative wave-function profiles in  Figs.~\ref{Fig7}(i)--\ref{Fig7}(l) further support the coexistence of fractal-like and localized states in this regime.

These results suggest four localization regimes in the $(\lambda,t_1)$ plane for $n=2$ in the $\beta\to0$ limit: an extended regime, an intermediate regime with mobility-edge-like behavior, a mixed regime containing fractal-like and localized states, and a localized regime. Thus, mobility-edge-like and fractal-state structures may coexist with the reentrant topological windows in the smooth integer-power quasiperiodic limit.

\subsection{Bulk localization properties in other parameter regimes}

\begin{figure}
\centering
\includegraphics[width=0.65\linewidth]{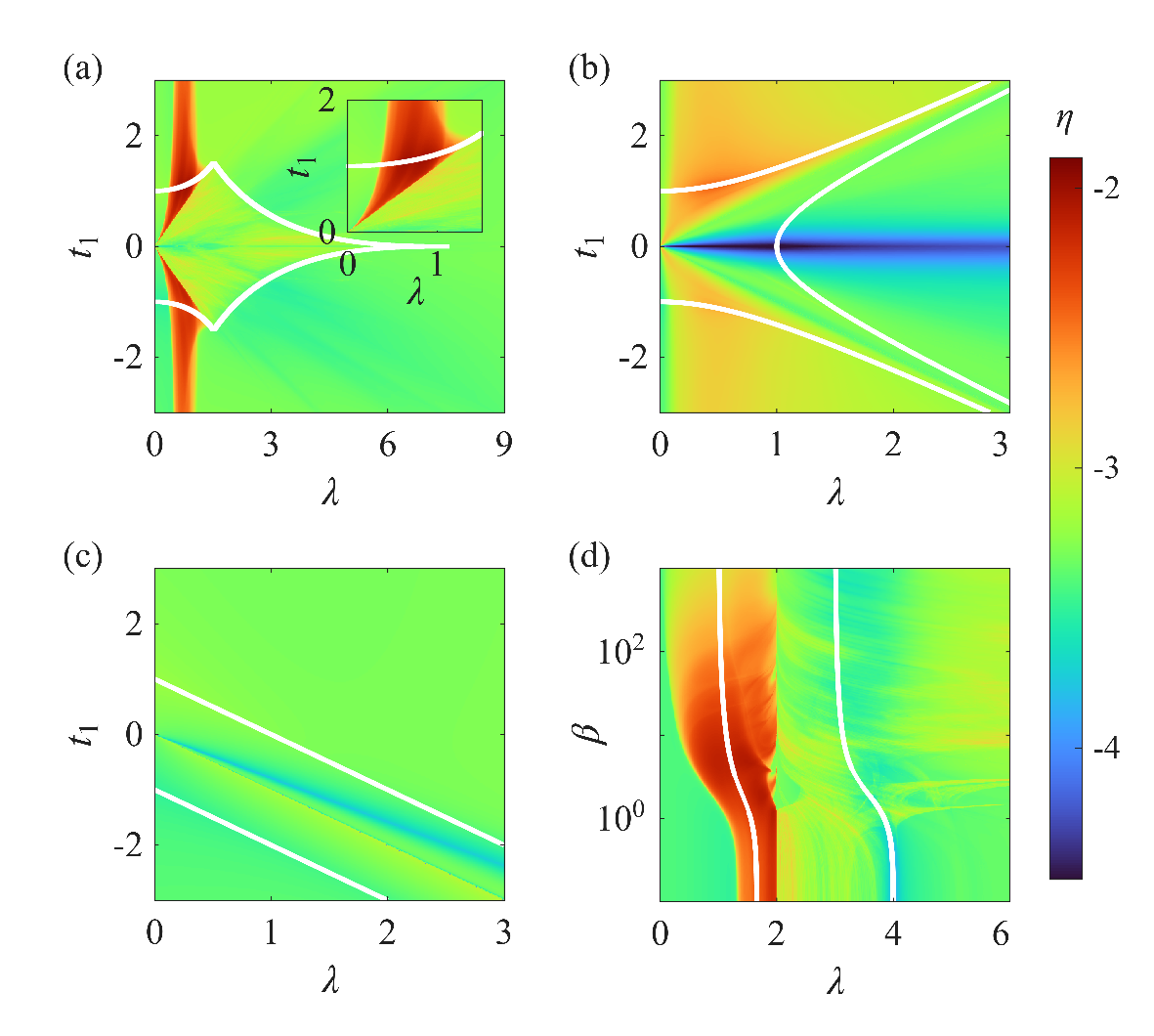}
\caption{\label{Fig8}Bulk localization properties in other parameter regimes. The color maps show the diagnostic quantity $\eta$ as a function of $(t_1,\lambda)$ in (a)--(c), and as a function of $(\beta,\lambda)$ in (d). The white solid curves denote the topological phase boundaries determined by $\nu=0$. (a) $\beta\to0$ and $n=3$. (b) $\beta\to\infty$ and $n=3$. (c) $\beta\to\infty$ and $n=2$. (d) Finite-$\beta$ interpolation for $n=2$ and $t_1=-2$. Here $N=610$ unit cells, corresponding to $L=1220$ lattice sites, are used. When boundary conditions are required, PBC are used.}
\end{figure}

We apply the same analysis to the $n=3$ case in the $\beta\to0$ limit and find a similar four-regime structure in the bulk localization properties, as shown in Fig.~\ref{Fig8}(a). The endpoints of the enhanced-$\eta$ regions are again visibly correlated with the vertices of the topological phase boundaries. We have also checked the $n=4$ case numerically and found the same qualitative trend. These results suggest that, in the smooth integer-power quasiperiodic limit, changes in the bulk localization properties are correlated with parts of the topological phase boundaries.

For the $\beta\to\infty$ limit, the situation is different. Figs.~~\ref{Fig8}(b) and \ref{Fig8}(c) show the corresponding $\eta$ maps for $n=3$ and $n=2$, respectively. In contrast to the $\beta\to0$ case, we do not find clear boundaries separating the four localization regimes discussed above. Moreover, the correlation between the enhanced-$\eta$ regions and the topological phase boundaries becomes much weaker. This is consistent with the fact that, in the sign-function limit, the modulation profile is square-wave-like for odd $n$ and effectively uniform for even $n$.

Finally, Fig.~\ref{Fig8}(d) illustrates the finite-$\beta$ interpolation for $n=2$ at fixed $t_1=-2$. As $\beta$ increases, the enhanced-$\eta$ region gradually weakens and its boundary becomes less sharp, indicating that the distinction between different bulk-localization regimes becomes progressively less pronounced. Therefore, while mobility-edge-like and fractal-state features can coexist with the reentrant topological windows in the smooth quasiperiodic limit, the topological phase boundaries discussed in the main text are governed by the zero-energy transfer criterion $\nu=0$ and do not require a simultaneous localization transition of all bulk states.

\section{Robustness of the topological phase diagram}\label{sec:robustness}

In Section \ref{sec:experiment}, we propose an electrical-circuit realization of the SSH model. In realistic circuits, however, capacitors and resistors inevitably have manufacturing tolerances, and active elements such as operational amplifiers have finite bandwidths. These imperfections may introduce random fluctuations into the effective hopping amplitudes. We therefore examine whether the predicted even--odd phase diagrams remain robust against such nonidealities.

\begin{figure}
\centering
\includegraphics[width=0.7\linewidth]{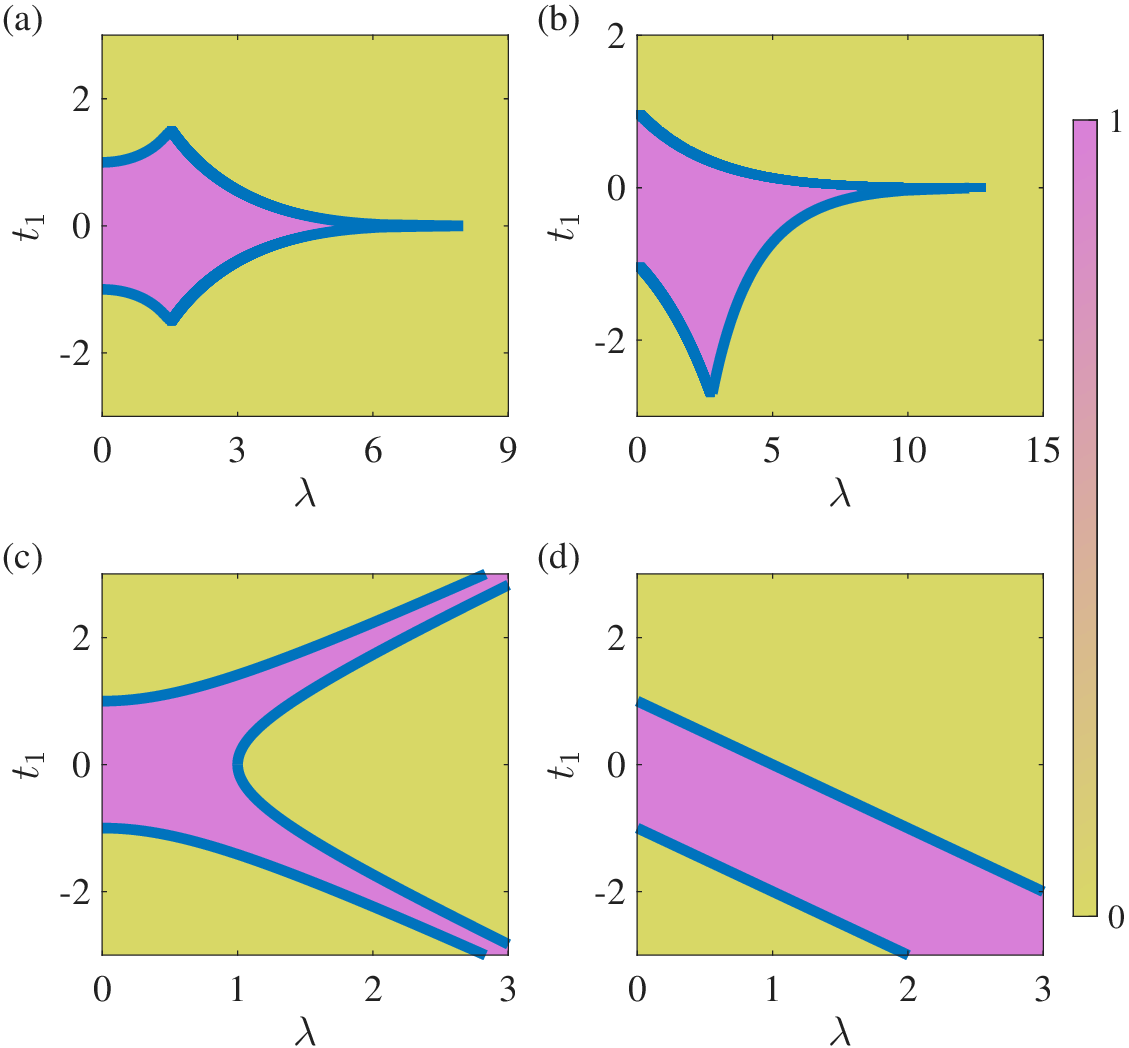}
\caption{\label{Fig9}Robustness of the topological phase diagrams against random hopping fluctuations. (a) $\beta\to0$, $n=3$. (b) $\beta\to0$, $n=4$. (c) $\beta\to\infty$, odd-$n$ case represented by $n=1$. (d) $\beta\to\infty$, even-$n$ case represented by $n=2$. The white solid curves denote the analytical phase boundaries of the clean system. In evaluating the topological indicator in Eq.~(\ref{eq4}), random hopping fluctuations with amplitudes up to 5\% are added to both intracell and intercell hoppings. The plotted results are averaged over 100 random configurations. Here $N=610$ unit cells, corresponding to $L=1220$ lattice sites, are used.}
\end{figure}

To model component tolerances, we introduce multiplicative random fluctuations in the hopping amplitudes,
\begin{equation}
\omega_j\rightarrow \omega_j(1+\eta_j),\qquad
t_2\rightarrow t_2(1+\xi_j),
\end{equation}
where $\eta_j$ and $\xi_j$ are independent random variables uniformly distributed in $[-0.05,0.05]$ for intracell and intercell hoppings, respectively. The topological indicator is then evaluated as
\begin{equation}
Q=\frac{1}{2}\left[1-\mathrm{sgn}\left(\prod_j \omega_j^2(1+\eta_j)^2-\prod_j t_2^2(1+\xi_j)^2\right)\right].
\end{equation}
For each parameter point, $Q$ is averaged over 100 random configurations. This model captures the dominant effect of capacitance tolerances on the off-diagonal hopping amplitudes. Frequency-dependent corrections caused by finite operational-amplifier bandwidths are expected mainly to broaden the impedance resonances, provided that the operating frequency remains well within the bandwidth of the active elements. These corrections are not expected to alter the topological regions away from the phase boundaries.

The averaged $Q$ results are shown in Fig.~\ref{Fig9}. The main structure of the phase diagrams remains unchanged under 5\% random hopping fluctuations, and the averaged numerical boundaries agree well with the analytical transition lines of the clean system. Small deviations and broadening occur mainly near the phase boundaries, where the bulk gap or zero-mode inverse localization length becomes small. Therefore, in realistic circuits, the ideal sharp transition lines should be interpreted as narrow crossover regions, while the even--odd structure and the reentrant topological windows remain observable away from the critical boundaries.


\bibliographystyle{cas-model2-names}




\end{document}